\documentclass[12pt]{iopart}
\usepackage{iopams}
\usepackage{amssymb,graphicx,epsfig,multirow}
\begin{document}

\title{Evolution of structure, magnetic and transport properties of Fe$_{1-x}$Mn$_{x}$Se}
\author{Kapil E. Ingle$^1$, K. R. Priolkar$^1$, P. A. Bhobe$^2$, A. K. Nigam$^3$}
\address{$^1$Department of Physics, Goa University, Taleigao Plateau, Goa 403 206 India}
\address{$^2$Indian Institute of Technology Indore, Madhya Pradesh 453331 India}
\address{$^3$Tata Institute of Fundamental Research, Dr. Homi Bhabha Road, Colaba, Mumbai 400005, India}
\ead{krp@unigoa.ac.in}

\date{\today}

\begin{abstract}
The present paper seeks to investigate effect of Mn doping in superconducting FeSe. It is found that over the entire doping range in Fe$_{1-x}$Mn$_x$Se ($0 \le x \le 1$), Mn does not substitute Fe in the superconducting tetragonal phase. Instead two impurity phases, NiAs type hexagonal phase and NaCl type cubic phase grow with increasing Mn content. Initially, hexagonal phase has a higher content than the cubic phase but beyond $x$ = 0.5, the cubic phase grows rapidly and for $x \ge 0.8$, the sample is monophasic with cubic NaCl type structure. The superconducting tetragonal phase content steadily decreases with increasing Mn concentration and completely disappears beyond $x = 0.5$. The premise that Mn never replaces Fe in the superconducting phase is further strengthened by observation of a sharp drop in AC susceptibility akin to superconducting transition at the T$_c$ of FeSe up to $x$ = 0.5. EXAFS studies at the Fe K edge also show that the Fe has a four coordinated tetragonal local structure in all compositions below $x = 0.5$,  similar to that in FeSe and it gradually changes to a six coordinated one as is expected for a NaCl type cubic phase for $x \ge 0.5$.
\end{abstract}

\vspace{2pc}

%\submitto{\MRX}

\maketitle

\section{Introduction}
Substitutions play an important role in deciding the properties and ground state of superconducting compounds. For example, substitutions help to understand the symmetry of superconducting state \cite{bal,liu} or to identify topological superconductors \cite{hu} as well as to improve the properties such as to pin vortices and increase the critical current density of superconductors \cite{tin}. Superconductivity, especially in cuprates and Fe based compounds, can also be induced through substitutions \cite{bed,mizu2,traq}.

Effects of substitution are varied and depend upon which of the atom/ion in the parent compound is being substituted. For example, substitution of the rare-earth ion by alkaline earth ions like Ba or Sr in La$_2$CuO$_4$ induces superconductivity \cite{bed} but any substitution of either magnetic or non magnetic ion at the Cu site results in suppression of superconductivity \cite{xiao}. Similarly in iron based superconductors, superconductivity can be induced in FeTe by  replacing Te by Se \cite{ymizu,mhfang} or by S \cite{mizu2,si} or even oxygen annealing\cite{sun}. However, substitutions at the Fe site result in destruction of superconducting state of FeSe \cite{zangam,huang,awana}. This scenario is also valid in case of other families of Fe based superconductors \cite{tan,yy} though there are notable exceptions like \cite{zz}.

In case of iron chalcogenides, substitution of Fe by other transition metals except Mn results in drastic suppression of $T_c$ \cite{mm}. Thought the main reason of this suppression of superconductivity is considered to be magnetic pair breaking, recent single crystal studies have shown that other processes need to be considered to be explain the suppression in transition metal doped iron chalcogenide superconductors \cite{yu}. The case of Mn substitution however is different. It has been argued that while other transition metal atoms replace Fe in the superconducting phase, Mn tends to induce an additional impurity phase \cite{mm}. Recent studies on transition metal doping in FeSe reveal that a mere  2 to 5\% Ni substitution in Fe$_{1-x}$Ni$_x$Se completely suppresses the superconducting state but a similar amount of Mn doping leaves the superconducting state unaffected \cite{kapil3}. What happens to the superconducting state at higher doping concentrations? How do the two structural phases evolve?

Secondly, despite structural similarities in form of stacked square layers of transition metal atoms in the iron chalcogenide and cuprate superconductors, the non superconducting ``parent'' compounds have contrasting physical properties. While La$_2$CuO$_4$, for instance, is an antiferromagnetic insulator \cite{var}, FeTe possesses a metallic antiferromagnetic ground state believed to be due to conduction carrier mediated spin density wave \cite{awana2015}. This is also true for other families of Fe based superconductors like Ba$_{1-x}$K$_x$Fe$_2$As$_2$ where the antiferromagnetic ordering of the parent compound ($x$ = 0) can only be explained on the basis of itinerant magnetism models \cite{john2}. Therefore, isostructural BaMn$_2$As$_2$ which is an antiferromagnetic insulator is proposed as the bridge between the layered iron pnictides and cuprates\cite{pandeyPRL}. Is there a similar bridging compound for iron chalcogenide superconductors? MnSe though an antiferromagnetic insulator, crystallizes in NaCl type cubic structure \cite{jbc} while the superconducting  phase of FeSe has  a tetragonal structure. It is therefore necessary to study the effect of Mn substitution in FeSe over the entire concentration range (a) to understand the evolution of the hexagonal impurity phase and its effect on superconductivity of FeSe and (b) to search for the existence of a bridging compound like BaMn$_2$As$_2$ in the Mn substituted FeSe.

\section{Experimental}
Fe$_{1-x}$Mn$_{x}$Se ($x$ = 0, 0.02, 0.05, 0.1, 0.2, 0.3, 0.4, 0.5, 0.6, 0.7, 0.8, 0.85, 0.9, 0.95 and 0.98) were prepared by solid state reaction method. All the required elements of  high purity Fe (99.98\%), Mn (99.9\%) and Se (99.99\%) were weighed as per stoichiometric ratio calculations. These elements were then ground to a fine powder, thoroughly mixed and pressed into 8 mm pellets under a uni-axial pressure of 5 tons. These pellets were then vacuum sealed in a quartz ampoules under an vacuum of $\sim 10^{-5}$ Torr. These vacuum-sealed samples were heat-treated through different steps to $650^\circ$ and dwelled for 24 hours. The samples were then furnace cooled to room temperature. X-Ray Diffraction measurements were performed on  these samples using Rigaku X-ray diffractometer  with Cu K$_\alpha$ ($\lambda$ = 1.5418\AA) radiation in the scattering angular (2$\theta$) range of $10^\circ $ to $80^\circ $ in equal 2$\theta$ steps of 0.02$^\circ$ for phase identification.  The diffraction patterns were Rietveld refined using FullProf Suite \cite{rodgir} to obtain lattice parameter variation  and variation of phase fraction as a function of Mn concentration in FeSe. Electrical resistivity of all the samples was measured using the standard four probe method in the temperature range 10K - 300K. DC and AC magnetization measurements were carried out using a Quantum Design MPMS SQUID magnetometer and ACMS option on Physical Property measurement system (PPMS) respectively. X-ray absorption fine structure measurements at the Fe K edge were performed in transmission using the R-XAS Laboratory EXAFS spectrometer. The generator was operated at 14 kV and 40 mA. A Ge(220) crystal was used as a monochromator. The incident (I0) intensities were measured using an ionization chamber filled with Ar gas while a scintillation detector was used for measuring transmitted (I) intensities. Absorber thickness was appropriately adjusted to restrict the absorption edge jump ($\Delta\mu$) to an optimum value. The edge energy was calibrated using Fe metal foil as standard. The EXAFS data was reduced following standard procedures in Demeter program \cite{ravel}.

\section{Results and Discussion}
\begin{figure}
\centering
\includegraphics[width=\columnwidth]{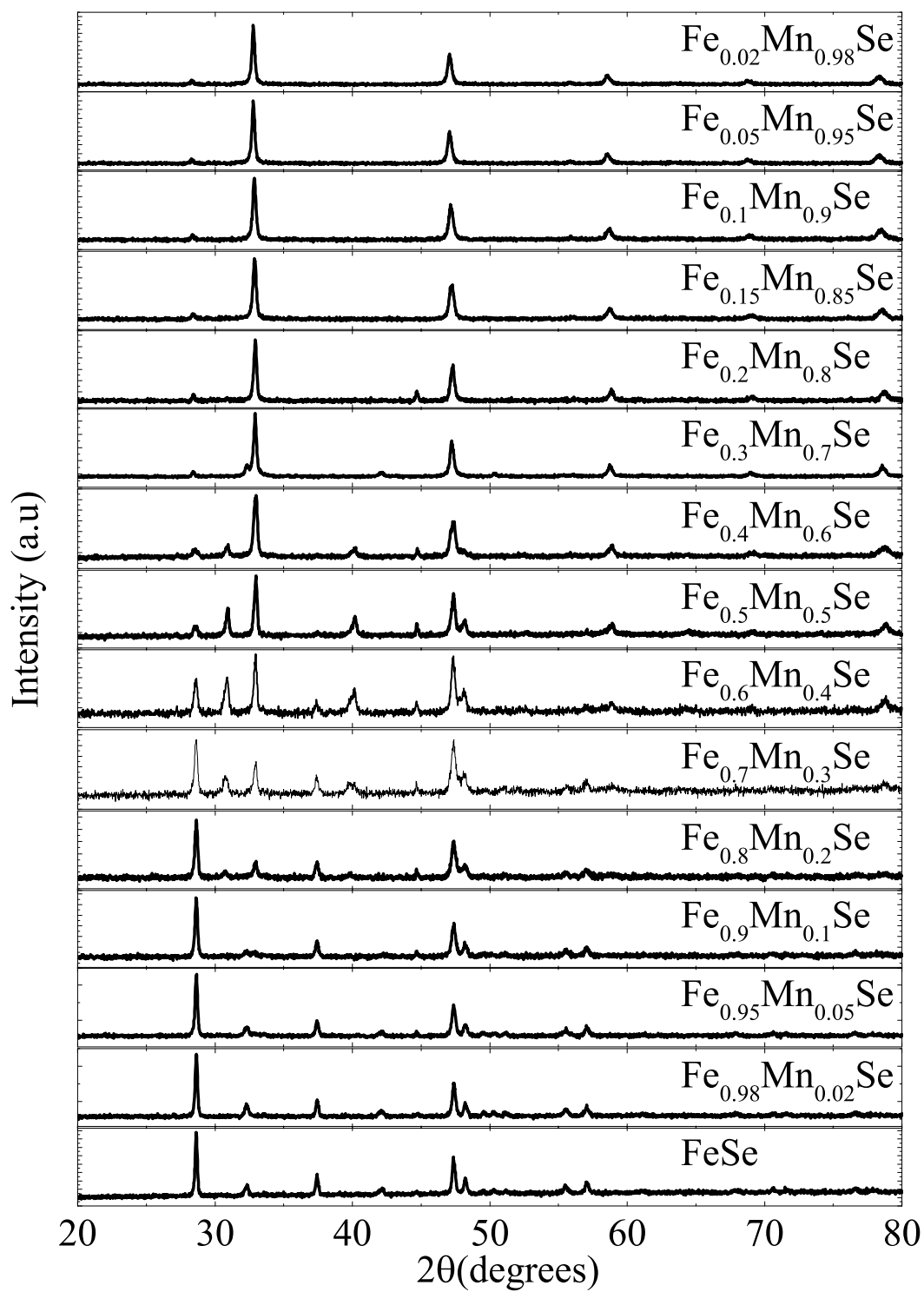}
\caption{\label{figure1} Room temperature XRD patterns of Fe$_{1-x}$Mn$_{x}$Se (x=0, 0.02, 0.05, 0.1, 0.2, 0.3, 0.4, 0.5, 0.6, 0.7, 0.8, 0.85, 0.9, 0.95).}
\end{figure}

X-ray diffraction patterns of all the Fe$_{1-x}$Mn$_x$Se samples are presented in Figure \ref{figure1}. FeSe ($x$ = 0) crystallizes in a tetragonal structure  (Sp. gr. P4/nmm) along with a small quantity ($\sim$ 1\%) of impurity  hexagonal NiAs type phase. While the Mn rich ($x$ = 0.98) compound exhibits a diffraction pattern corresponding to a pure cubic NaCl type structure. As noted above MnSe has a cubic structure at room temperature.

\begin{figure}
\centering
\includegraphics[width=\columnwidth]{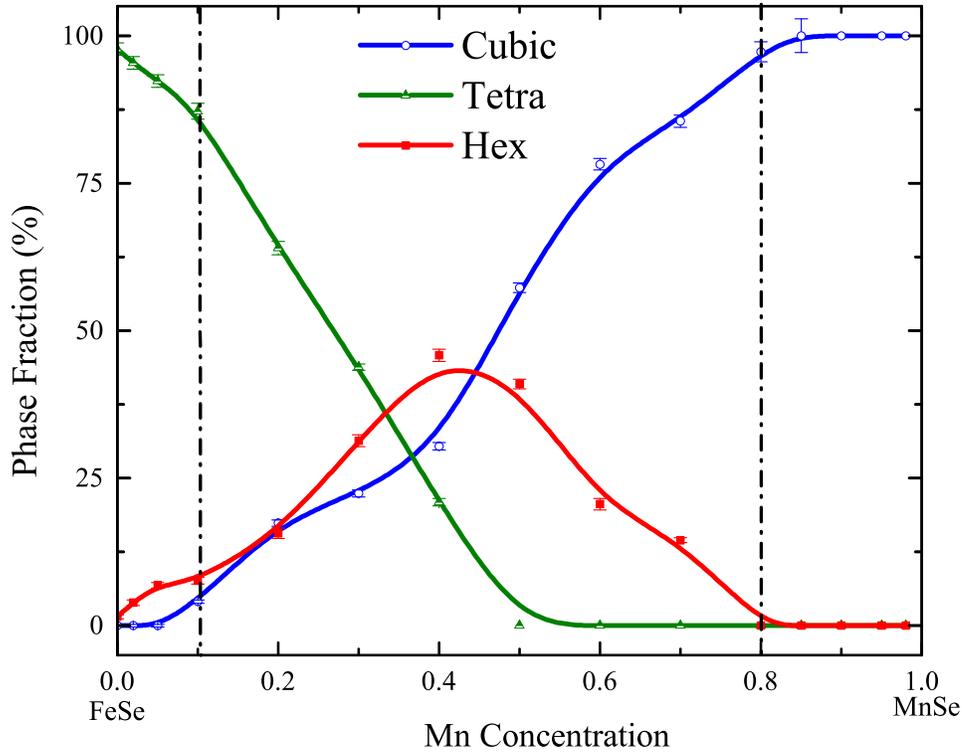}
\caption{\label{figure2} Variation of tetragonal, hexagonal and cubic phase fractions in Fe$_{1-x}$Mn$_{x}$Se as a function of Mn concentration.}
\end{figure}
\begin{figure}
\centering
\includegraphics[width=\columnwidth]{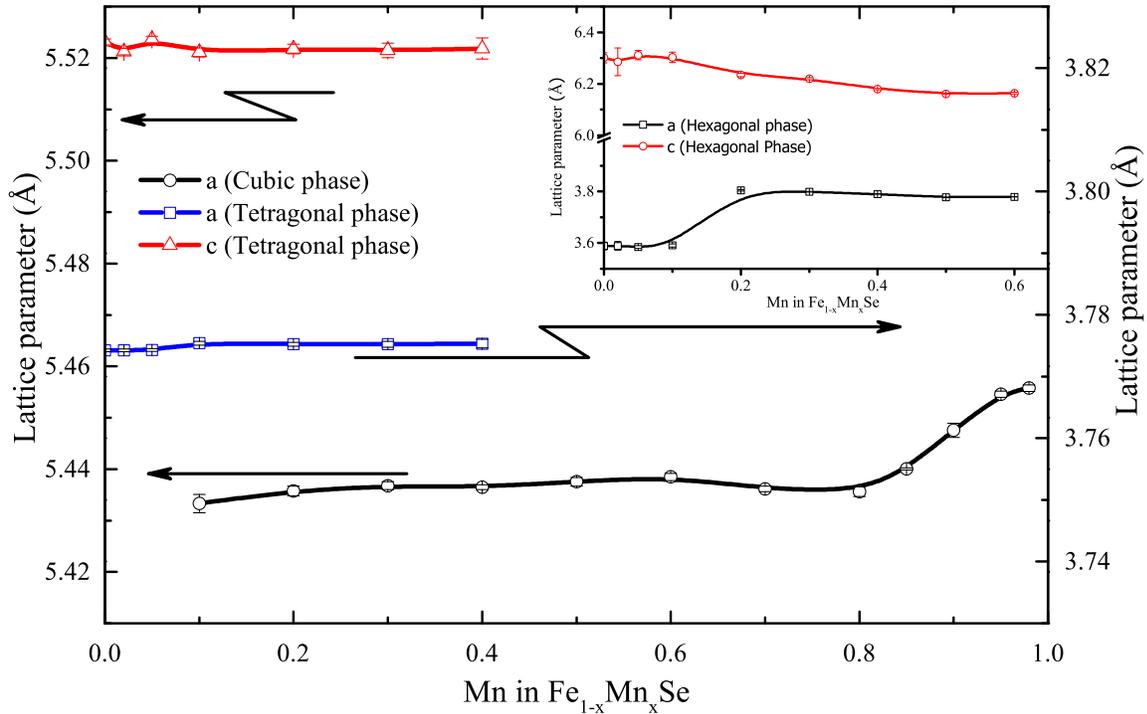}
\caption{\label{figure3} Variation of lattice parameter parameters of tetragonal and cubic phases as function of Mn concentration in Fe$_{1-x}$Mn$_{x}$Se.}
\end{figure}

In order to understand the variation of different phase fractions as a function of Mn concentration, the diffraction patterns were Rietveld refined. The obtained variation of phase fractions is plotted as a function of Mn concentration in Figure \ref{figure2}. It can be seen that up to $x$ = 0.05, the compounds are essentially biphasic consisting of major ($>$ 90\%) tetragonal phase and minor hexagonal (NiAs type) phase. Recently it was shown that in these compounds, the dopant, Mn preferentially substitutes Fe in hexagonal phase leaving the superconducting tetragonal phase unaffected \cite{kapil3}. From $x$ = 0.1, a third phase, cubic (NaCl type) appears and its concentration grows almost linearly with increasing Mn concentration until it reaches 100\% for $x \ge 0.8$. The fraction of the hexagonal phase also grows up to $x$ = 0.4 and then decreases to zero at $x$ = 0.8. Along with this variation in phase fractions of cubic and hexagonal phases, the fraction of tetragonal phase decreases continuously to zero at $x \approx 0.5$. Such a variation in phase fractions, especially the increase in cubic phase fraction concomitant with decrease in tetragonal phase fraction with increasing $x$ gives weight to the argument that Mn does not replace Fe in the tetragonal phase of FeSe, instead separates out in an impurity phase. This argument is further supported by the variation of lattice parameters which is depicted in Figure \ref{figure3}. Over the entire concentration range where the tetragonal phase exists, both lattice constants $a_{tetra}$ and $c_{tetra}$, show very little variation in their values. The same is also true in case of variation of lattice constant of the cubic phase ($a_{cubic}$). Except for a small increase in its value beyond $x = 0.8$, $a_{cubic}$ too remains largely constant over the rest of the concentration range. The increase at higher values of $x$ could be understood to be due to substitution of Fe for Mn in the cubic phase. It may be noted that for $x \ge 0.8$, the compounds are 100\% cubic in structure and therefore substitution of Fe for Mn in cubic phase is but natural. Also the fact that Fe has a smaller size as compared to Mn results in decrease in $a_{cubic}$ with decreasing Mn concentration ($x$). The variation of lattice parameters of the hexagonal phase are presented as inset to Figure \ref{figure3}. Here the lattice parameter $a_{hexa}$ gradually increases for smaller values of $x$, followed by a relatively steeper increase in its value between $0.1 \le x \le 0.2$ and then again remaining constant for higher values of $x$. Pure FeSe contains about 1\% hexagonal impurity. Therefore the increase in $a_{hexa}$ could be due to replacement of Fe by larger Mn and beyond $x$ = 0.1, the hexagonal phase could be largely composed of Mn atoms. This augers well also with the percentage of phase fractions. At $x$ = 0.1, the sum total of phase fraction of cubic and hexagonal phases is about 12\% indicating both these phases to be largely made of MnSe. The increase in $a_{hexa}$ up to $x = 0.2$ is accompanied by a continuous decrease in $c_{hexa}$. It may be noted that FeSe is superconducting and a larger $c$ parameter helps in localizing the Cooper pairs in basal planes while a lower value of $c$ in hexagonal MnSe aids in strengthening the antiferromagnetic coupling along $c$-axis \cite{jbc}. For $x > 0.4$, where the tetragonal phase fraction is nearly zero and the hexagonal phase fraction also decreases, Fe along with Mn crystallizes in cubic NaCl type phase. These facts point towards non substitution of Fe by Mn in the superconducting tetragonal phase.

\begin{figure}
\centering
\includegraphics[width=\columnwidth]{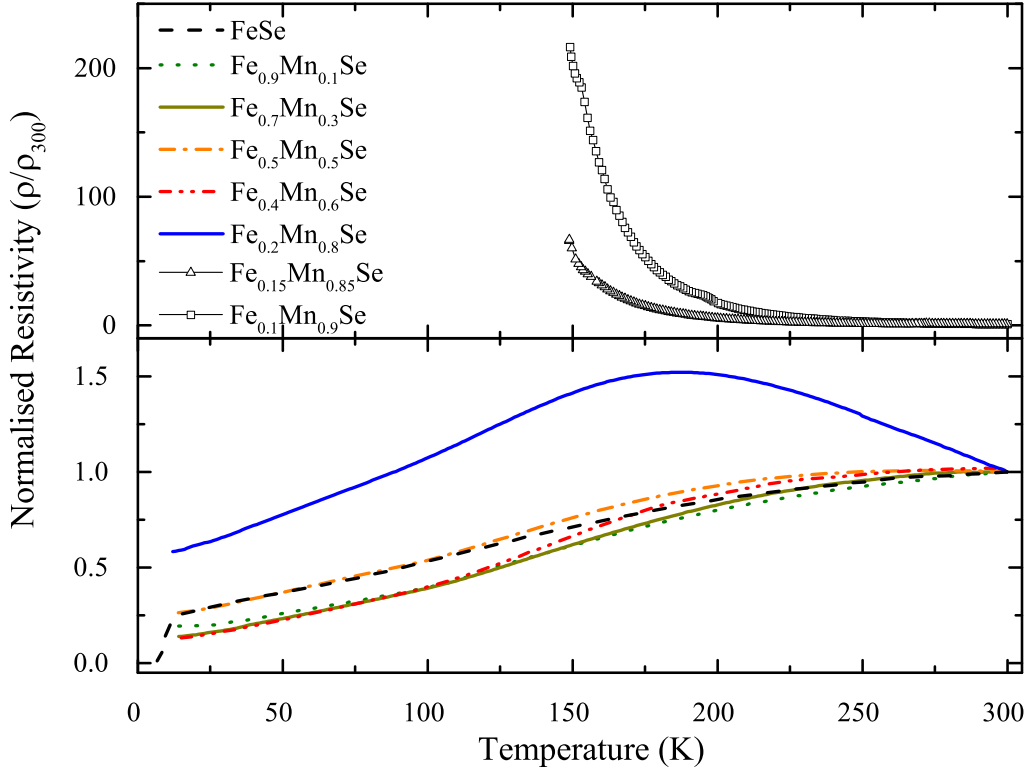}
\caption{\label{figure4} Variation of normalized resistivity ($\rho(T)/\rho(300 K)$) as function of temperature for Fe$_{1-x}$Mn$_{x}$Se}
\end{figure}

Normalized resistivity as a function of temperature for some of the compounds are presented in Figure \ref{figure4}. In Figure \ref{figure4}, all compounds except Mn rich samples ($x \ge 0.8$) exhibit metallic behavior over the entire measurement range. Superconducting transition at T${\rm c}$ = 8K can be seen in case of FeSe. Such a transition at nearly the same temperature was also seen in compounds with $x < 0.1$ (not shown in figure). Although metallic behavior was seen even in other samples with $x < 0.8$, superconducting transition was not recorded down to 5K in resistivity measurements. Since these samples are multiphasic, metallic resistivity behavior implies that there are at least a few interconnected metallic regions which allow the flow of charges across the sample. This is important because the fraction of tetragonal phase which is known to be metallic decreases rapidly and vanishes completely beyond $x$ = 0.4. Therefore in compounds where $0.4 < x < 0.8$, one of the two phases, hexagonal or cubic have to be metallic and such regions should be connected in some way to give a metallic resistivity behavior. Considering the fact that MnSe exhibits semiconducting resistivity \cite{jbc} and Fe$_{0.2}$Mn$_{0.8}$Se which has nearly 100\% cubic structure exhibits metal insulator transition (Figure \ref{figure4}(a)), it can be argued that more than 20\% Fe substitution in the cubic phase of MnSe results in metallic behavior of resistivity. This argument is further strengthened by the fact that all Fe$_{1-x}$Mn$_x$Se ($0.4 < x < 0.8$) have more than 20\% Fe content in the cubic phase. For instance, Fe$_{0.5}$Mn$_{0.5}$Se has about 58\% cubic phase and about 42\% hexagonal phase. If one assumes random substitution of Fe and Mn in both the phases then about 30\% of Fe content will be in the cubic phase. Compounds with less than 20\% Fe show semiconducting resistivity behavior as can be seen from Figure \ref{figure4}(b).

The above structural studies indicate that Mn does not substitute Fe in tetragonal phase but crystalizes in hexagonal and cubic phases. Resistivity behavior indicates presence of Fe in cubic phase. Therefore the questions arise as to what happens to the superconducting transition in Mn doped FeSe? What is the magnetic ground state of these compounds? And is there any correlation between the structure, resistivity and magnetic properties of Fe$_{1-x}$Mn$_x$Se? To seek answers to these questions DC magnetization measurements have been performed on some of these samples. The results are presented in Figure \ref{figure5} and further details are presented in Supplementary (Figure S1). It can be seen from Figure \ref{figure5}(a) that magnetization acquires negative values below 8K corresponding to superconducting transition in FeSe. At around the same temperature, the 30\% and 50\% Mn doped FeSe samples also exhibit a drop in magnetization value indicating the onset of superconducting transition. The positive value of  magnetization and increase in its magnitude with increasing Mn concentration can be ascribed to magnetic contributions from other structural phases present in these compounds. Fe$_{0.7}$Mn$_{0.3}$Se has significant presence of tetragonal phase and the sudden drop in magnetization around 7K strengthens the argument that Mn does not substitute Fe in tetragonal phase. A similar drop in magnetization, though much weaker is also seen in $x$ = 0.5 wherein the tetragonal phase has nearly disappeared. The similarities in magnetization behavior of FeSe, Fe$_{0.7}$Mn$_{0.3}$Se and Fe$_{0.5}$Mn$_{0.5}$Se indicate the role of local structural correlations in superconducting transition in FeSe \cite{kapil2}.

\begin{figure}
\centering
\includegraphics[width=\columnwidth]{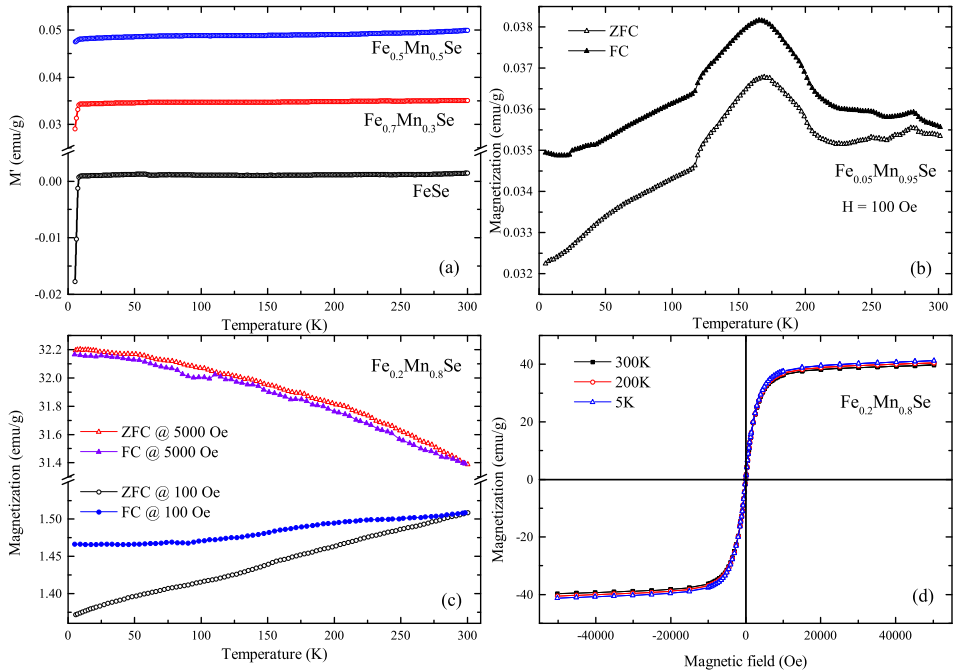}
\caption{\label{figure5} (a) Temperature dependence of zero field cooled (ZFC) ac magnetization at H=10Oe for FeSe, Mn$_{0.3}$Fe$_{0.7}$Se and Mn$_{0.5}$Fe$_{0.5}$Se, (b) Magnetization versus temperature behavior for Mn$_{0.95}$Fe$_{0.05}$Se recorded in applied field of 100 Oe during ZFC and field cooled (FC) cycles, (c) DC magnetization measurements of Mn$_{0.8}$Fe$_{0.2}$Se in 100 Oe and 5000 Oe recorded during ZFC and FC cycles, (d) Isothermal magnetization (M(H)) for Mn$_{0.8}$Fe$_{0.2}$Se recorded in applied field of $\pm$ 50000 Oe at 300K, 200K and 5K.}
\end{figure}

\begin{figure}
\centering
\includegraphics[width=\columnwidth]{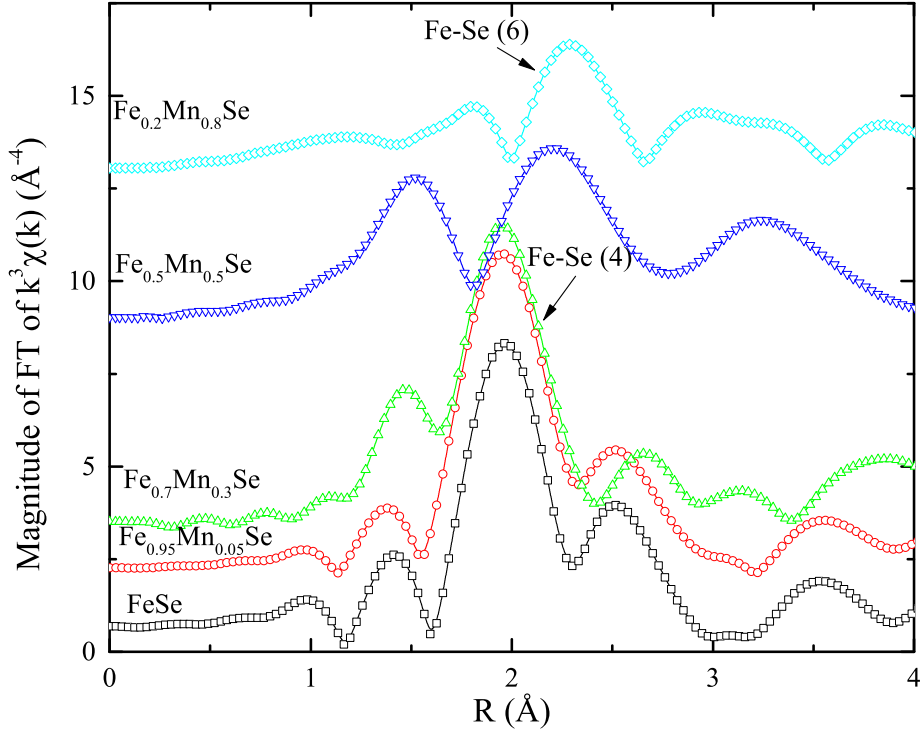}
\caption{\label{figure6} Magnitude of FT of Fe K edge EXAFS in Fe$_{1-x}$Mn$_x$Se, $x$ = 0, 0.05, 0.3, 0.5 and 0.8 compounds. Peaks indicated by arrows correspond to Fe-Se correlations with 4 nearest neighbors (Fe - Se (4)) and 6 nearest neighbors (Fe - Se(6)). }
\end{figure}

In order to confirm this, we have performed Fe K edge EXAFS on five compositions ($x$ = 0, 0.05, 0.3, 0.5 and 0.8). The magnitude of Fourier transform (FT) of EXAFS signal is presented in Figure \ref{figure6}. In tetragonal phase, the Fe is tetrahedrally coordinated to Se with a bond length of 2.38\AA~ while in the cubic phase the Fe-Se bond distance is 2.7\AA~ and has octahedral coordination. It can be observed in Figure \ref{figure6} that the first peak in the magnitude of FT of EXAFS corresponding to Fe-Se correlation shifts from $\sim 2$\AA~ to $\sim 2.3$\AA~ as the phase fraction of tetragonal phase decreases from nearly 100\% in $x$ = 0 to 0\% in $x \ge  0.5$. The coordination number of Fe also changes from 4 (tetrahedral) to $\sim$6 (octahedral). A contribution of 4 coordinated Fe to EXAFS signal can be noticed up to $x$ = 0.5 thus indicating that the sudden drop in magnetization akin to superconducting transition seen in Fe$_{0.7}$Mn$_{0.3}$Se and Fe$_{0.5}$Mn$_{0.5}$Se is due to  presence of tetrahedrally coordinated Fe in these compounds. It has been reported that superconductivity in iron chalcogenides is critically dependent on the tetrahedral Fe-Se correlation which defines the extent of Fe $3d$ and Se $4p$ hybridization. This hybridization is believed to be responsible for superconductivity \cite{kapil2}. The present EXAFS studies indicate that superconducting interactions persist so long as tetrahedrally coordinated Fe atoms are present in the compound.

Returning back to magnetization studies, it can be seen from Figure \ref{figure5}(b) that the Mn rich compound, Fe$_{0.05}$Mn$_{0.95}$Se has an antiferromagnetic ground state. The compound exhibits two antiferromagnetic transitions at about 275K and at 165K. This magnetization behavior is quite similar to that exhibited by MnSe wherein the first transition at about 270K occurs due to transformation of a certain volume fraction of the compound to antiferromagnetically ordered NiAs phase. The second transition at about 125K is attributed to the antiferromagnetic ordering of the remaining cubic phase \cite{jbc}. A similar behavior of magnetization is also seen in compound with $x$ = 0.85.

DC Magnetization curves for $x$ = 0.8 recorded in applied magnetic fields (H) of 100 Oe and 5000 Oe during ZFC and FC cycles are presented in Figure \ref{figure5}(c). A completely opposite behavior of the two magnetization curves is noted. While magnetization decreases with decrease in temperature when recorded in 100 Oe applied field, it shows an increase when recorded in higher field of 5000 Oe. This hints towards field induced magnetization. No magnetic transition is seen in the entire temperature range 5K - 300K, though small humps can be noted in the data recorded in H = 100 Oe. It is to be noted that FC magnetization recorded at 5000 Oe has a lower magnitude as compared to that of ZFC magnetization. Therefore magnetization curves were recorded in several fields between 100 Oe and 5000 Oe and these curves are presented in Supplementary figure S1. Effect of magnetic field in inducing magnetization can be seen as the applied magnetic field is increased from 100 Oe to 5000 Oe, but along with this increasing magnetization, the crossover between ZFC and FC magnetization is also clearly seen. This behavior implies presence of magnetic inhomogeneities.

Isothermal magnetization recorded in H = $\pm$ 50000 Oe at 300K, 200K and 5K and presented in Figure \ref{figure5}(d) are nearly identical in magnitude and shape. The `S' shaped loop with zero coercivity indicates presence of short range ferromagnetic correlations in the compound. The M(H) loops further strengthen the presence of magnetic inhomogeneities. Structurally this sample consists of only cubic chalcogenide phase and hence the magnetic behavior could be either due to presence of tiny amounts (not detected through x-ray diffraction) of other Mn(Fe)Se phases or due to magnetic phase separation of Mn rich and Fe rich chalcogenide regions. It is to be noted that onset of metallic resistivity and ferromagnetic correlations occur together. Hence the magnetic and transport properties could be correlated to each other. A more detailed study on magnetic and structural properties of these cubic Fe(Mn)Se systems is needed to arrive at proper understanding.

\section{Conclusions}
The above results indicate that path from superconducting FeSe to antiferromagnetic MnSe is quite complex. The structural phase diagram of Fe$_{1-x}$Mn$_x$Se solid solutions is quite rich with three structural phases. These structural phases arise because Mn does not replace Fe in the superconducting tetragonal phase. Instead it induces a hexagonal impurity phase and later an additional cubic NaCl type phase is formed. The phase fraction of the cubic phase grows continuously with increasing Mn concentration while the hexagonal phase disappears beyond $x$ = 0.8 after peaking at about $x$ = 0.5. The content of superconducting tetragonal phase on the other hand decreases continuously before disappearing at about $x$ = 0.4. A direct consequence of this is the preservation of superconducting interactions so long as the tetrahedral Fe - Se correlations are present. Thus the present study provides a clear reason for non suppression of superconductivity in Mn doped iron chalcogenides to be non substitution of Mn in the tetragonal FeSe phase. Due to this fact there is also no isostructural antiferromagnetic insulating phase in Mn doped FeSe which could act as a bridging compound between iron chalcogenide and cuprate superconductors.

\ack{Authors thank Department of Science and Technology, Govt. of India for financial assistance under the project SR/S2/CMP-57. Kapil Ingle acknowledges fellowship grants under UGC-BSR fellowship programme.}

\end{document}

\maketitle{\centerline {\em Supplementary material to the paper titled}
{\bf{\noindent Evolution of structure, magnetic and transport properties of Fe$_{1-x}$Mn$_{x}$Se}}\\

%\noindent{Author list}\\

%\noindent Corresponding author : krp@unigoa.ac.in \\

%\noindent This supplement contains \\
%\noindent Figure S1, Figure S2, Figure S3 Figure S4, Figure S5 and Figure S6\\

\begin{figure}[h]
\epsfxsize=150mm \epsffile{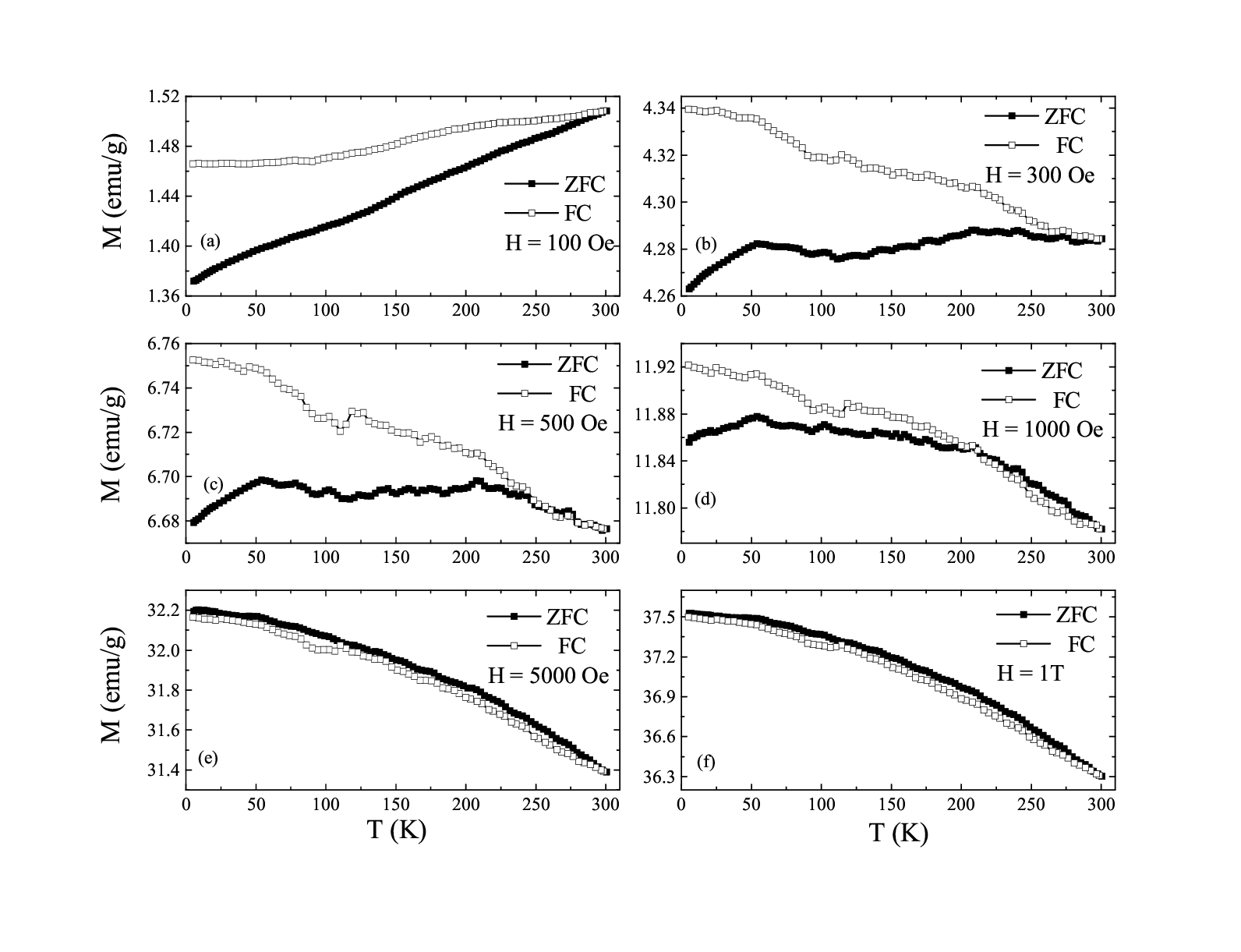}
\begin{center}
Figure S1: Temperature dependent DC magnetization of Fe$_{0.2}$Mn$_{0.8}$Se recorded during zero field cooled and field cooled cycles in different applied magnetic fields
\end{center}
\end{figure}

Figure S1 shows the temperature dependance of magnetization of Fe$_{0.2}$Mn$_{0.8}$Se recorded during zero field cooled and field cooled cycles in different applied magnetic fields. It can be seen that in 100 Oe the magnetization decreases with decreasing temperature. This behavior is reversed when the magnetic field is increased to 300 Oe. Though the magnetization value is quite appreciable, the curves appear noisy. These were repeated more than once to confirm the repeatability of the behavior. At about 1000 Oe, there is a crossover between ZFC and FC curves with ZFC magnetization having higher value than FC magnetization above 225K. For applied field of $\ge$ 5000 Oe ZFC magnetization has higher value than the FC magnetization over the entire temperature range.